\documentclass[a4paper,twoside]{article}

\usepackage{epsfig}
\usepackage{subcaption}
\usepackage{calc}
\usepackage{amssymb}
\usepackage{amstext}
\usepackage{amsmath}
\usepackage{amsthm}
\usepackage{multicol}
\usepackage{pslatex}
\usepackage[bottom]{footmisc}

\usepackage{booktabs}
\usepackage{multirow}
\usepackage{adjustbox}
\usepackage{natbib}

\usepackage{SCITEPRESS}     

\newcommand{\quotes}[1]{``#1''}

\begin{document}

\title{An Exploratory Study on the Evidence of Hackathons' Role in Solving OSS Newcomers' Challenges}

\author{\authorname{Ahmed Samir Imam Mahmoud\sup{1}\orcidAuthor{0000-0002-9898-2158}, Alexander Nolte\sup{1,2}\orcidAuthor{0000-0003-1255-824X} and Dietmar Pfahl\sup{1}\orcidAuthor{0000-0003-2400-501X}}
\affiliation{\sup{1}University of Tartu, Estonia}
\affiliation{\sup{2}Carnegie Mellon University, Pittsburgh, PA, USA}
\email{\{ahmed.imam.mahmoud, alexander.nolte, dietmar.pfahl\}@ut.ee}
}


\abstract{Background: OSS projects face various challenges. One major challenge is to onboard and integrate newcomers to the project.
Aim: We aim to understand and discuss the challenges newcomers face when joining an OSS project and present evidence on how hackathons can mitigate those challenges.
Method: We conducted two searches on digital libraries to (1) explore challenges faced by newcomers to join OSS projects, and (2) collect evidence on how hackathons were used to address them. We defined four evidence categories (positive, inconclusive, and no evidence) to classify evidence how hackathons address challenges. In addition, we investigated whether a hackathon event was related to an OSS project or not.
Result: We identified a range of newcomer challenges that were successfully addressed using hackathons. However, not all of the solutions we identified were applied in the context of OSS.
Conclusion: There seems to be potential in using hackathons to overcome newcomers' challenges in OSS projects and allow them to integrate faster into the project.}

\onecolumn \maketitle \normalsize \setcounter{footnote}{0} \vfill

\section{\uppercase{Introduction}}
\label{sec:intro}

Open source software (OSS) projects have proven their success through projects like Debian, Linux kernel, and many more. Researchers have studied their success factors~\citep{MATEOSGARCIA2008333,SADOWSKI2008323,1259211} and presented lessons learned~\citep{8632819} when creating and maintaining successful OSS projects. 

The management of OSS projects faces challenges. Maintaining engagement, attracting new developers~\citep{SEN2012364}, and integrating them such that they become contributors to a project are examples of such challenges. The range of challenges that exist in OSS projects has been discussed in literature~\citep{7985661,10.1145/2593728.2593732,NewcomersBarriers,10.1145/3125433.3125446}. Researchers have proposed and evaluated various approaches~\citep{7886910} that aim at helping the OSS communities to onboard newcomers and reduce the time it takes until they become productive.

Hackathons have previously been used to overcome challenges related to networking, sharing ideas, learning, and creating prototypes~\citep{Nolte2020HowTO}. Hackathons might thus have the potential to address problems related to onboarding newcomers to OSS projects and fostering their contribution. Hackathons are time-bounded events where individuals form ad-hoc teams and engage in intensive collaboration on a project idea of their interest~\citep{falk2022future}. These events have been organized and studied in various contexts, including corporations~\citep{pe2019designing,nolte2018you,komssi2015hackathons}, entrepreneurship~\citep{cobham2017appfest2,nolte2019touched}, and education~\citep{porras2019code,gama2018hackathon,kienzler2017learning}. Despite their widespread use, research focusing on hackathons in the context of OSS projects in general and on how they can aid the onboarding of newcomers, in particular, is scarce.

To address this gap, we conducted a review of existing literature on challenges that affect OSS newcomers. We contrasted our findings with prior work on how hackathons were used to tackle such challenges. In this position paper, we report on findings from this initial analysis. These findings will subsequently serve as a basis for an empirical study on how hackathons can support newcomers to join and become productive members of OSS projects.

We provide the following contribution: We present an overview of challenges affecting newcomers to start contributing to OSS projects reported in the literature. We also collect evidence about how hackathons were used to tackle these challenges. 

\section{\uppercase{Background}}
\label{sec:background}
%

\subsection{Challenges for OSS Projects}

OSS projects are an important piece of the larger software ecosystem contributing software, libraries, and packages. 
Typical challenges of OSS projects that need continuous attention and monitoring include the risk of underproduction~\citep{9426043}, the difficulty 
to attract and maintain developers~\citep{SEN2012364},
 knowledge management (KM)~\citep{dai_boujut_pourroy_marin_2020}, 
 and the handling of community dynamics 
 in agile OSS projects~\citep{8632819}.

In this paper, we focus on challenges related to onboarding newcomers to OSS projects and how hackathons were used to overcome them. 

\subsection{Developers Joining OSS Projects}

Prior research has been conducted to understand the onboarding of new developers joining OSS projects. This work includes understanding their motivation join OSS projects~\citep{10.5555/776816.776867}. Based on this understanding, researchers have proposed scripts for new developers to start contributing~\citep{VONKROGH20031217}, proposed a joining model~\citep{2049}, and proposed an approach to identify and recommend mentors in OSS projects~\citep{10.1145/2393596.2393647}. In addition, there is also work that aims to understand the impact of globalization and offshoring in OSS projects and provide tools to facilitate newcomers' learning~\citep{10.1145/1882362.1882452}.




\subsection{Hackathons and OSS}
It is a common practice for OSS projects to arrange in-person events like conferences or hackathons. Research on such events has found that they can aid the development of trust and build relationships~\citep{10.1145/3449249}. There are empirical studies on the Google Summer of Code (GSoC) -- a community code engagement event similar to hackathons. Findings indicate that such events can be used to attract new developers~\citep{10.1145/3131151.3131156}. Similarly, there are studies on how project characteristics can affect the onboarding of developers by analyzing data from a kick-start hackathon at Facebook. Findings indicate positive effects of mentoring on the onboarding process~\citep{10.1145/2652524.2652540}.
  
\section{\uppercase{Methodology}}
\label{sec:method}
Our work is divided into two main steps. First, we conducted a literature review on the challenges and barriers affecting newcomers in OSS projects and categorized them into different groups (section \ref{sec:method:sub:oss_challenges}). Second, we extracted evidence from existing work about how hackathons were used to overcome these challenges and barriers. 

\subsection{Identifying Challenges That Affect Newcomers in OSS Projects}
\label{sec:method:sub:oss_challenges}

We first searched for secondary studies on newcomer barriers in OSS projects, i.e., Systematic Literature Reviews (SLRs) in the ACM digital library, Google scholar, and IEEE Xplore. As keywords we used \textit{OSS}, \textit{newcomers}, \textit{barriers}, \textit{challenges}, \textit{SLR}, and \textit{systematic literature review}. One of the identified papers~\citep{STEINMACHER201567} specifically discussed barriers faced by newcomers in OSS projects and grouped them into different categories and subcategories.


Since SLRs do not cover the most recent studies, we expanded our search on the same digital libraries for the time period after the publication of the aforementioned paper~\citep{STEINMACHER201567} excluding the keywords \textit{SLR} and \textit{systematic literature review}. The goal was to identify additional challenges and barriers that were discussed after the SLR was published. 
This new search yielded 7 additional papers discussing newcomers' challenges and barriers in OSS projects. We only included papers that discussed developers joining OSS projects and excluded papers discussing newcomers in other settings.

From the final list of papers, we extracted all newcomers' challenges and barriers in OSS projects and classified them based on the categories proposed in the aforementioned SLR~\citep{STEINMACHER201567}. Because we found new types of barriers and challenges, we had to extend the original set of categories.


\subsection{Finding Evidence on Hackathons Addressing Newcomer Challenges}
\label{sec:method:sub:hackathons_evidence}


Before searching for prior work on how hackathons were used to overcome newcomer challenges, we defined categories that describe the state of evidence based on two criteria. The first criterion covers evidence that hackathons helped address newcomer challenges. The second criterion covers the context of the hackathon events, i.e., if they were related to OSS or not. Based on these criteria, we defined four categories. \textit{Positive evidence in open source} indicates that there is positive evidence in literature that hackathons have helped to overcome a specific OSS challenge. \textit{Positive evidence in different context} indicates there is positive evidence in literature that hackathons helped overcome a specific challenge outside of OSS. \textit{Inconclusive evidence} indicates that there is contradicting evidence in literature, i.e., that there is evidence in favor and against hackathons helping to overcome a specific challenge. \textit{No evidence} indicates a lack of evidence that hackathons help overcome a specific challenge. Table~\ref{table:abbreviations} provides an overview.

\begin{table}[htp]
\caption{Categories of evidence (including abbreviations)}
\label{table:abbreviations}
\centering
\begin{tabular}{@{}ll@{}}
\toprule
Evidence category             & Abbreviation \\ \midrule
Positive Evidence in OSS Projects      & P-E-OSS               \\ \midrule
Positive Evidence in Different Context & P-E-OTH               \\ \midrule
Inconclusive Evidence                  & INC-E                 \\ \midrule
No Evidence                            & NO-E                  \\ \bottomrule
\end{tabular}
\end{table}

After defining these categories, we conducted a search in digital libraries, including the ACM digital library, Google scholar, and IEEE Xplore, for papers that discuss newcomers and hackathon outcomes in general. There are no secondary studies on the topic so we used the following keywords in our search: \textit{hackathon}, \textit{newcomers}, \textit{outcome}, \textit{challenges}, \textit{engagement}, and \textit{collaboration}. In addition, we included newcomer challenges collected in the first step as keywords in our search.

Based on our search, we identified 8 papers. We analyzed these papers to find evidence that hackathons overcome or solve any of the previously identified challenges and barriers in the context of OSS and beyond. We focused on the method and approach described in the paper before extracting how newcomer challenges were addressed.

\section{\uppercase{Results}}
\label{sec:results}
Our main goal was to find evidence that hackathons helped overcome newcomer challenges in OSS projects. To achieve this goal, we collected and analyzed existing literature as discussed in~\ref{sec:method}. In this section, we present our findings regarding newcomer challenges (section~\ref{subsec:results:challenges}) and regarding evidence about how hackathons helped overcome them (section~\ref{subsec:results:evidence}).

\subsection{Challenges Affecting Newcomers in OSS Projects}
\label{subsec:results:challenges}
Several publications report on barriers and challenges of newcomers joining an OSS project that can be categorized into five main groups~\citep{STEINMACHER201567}. These include \textit{Finding a way to start}, \textit{Technical hurdles}, \textit{Poorly documented code}, \textit{Newcomers' previous knowledge}, \textit{Social interaction}, and \textit{Individuals problems}. We extended these categories by including more recent work. Table~\ref{t:evidence} shows the extended set of categories of challenges faced by newcomers in OSS projects. It also includes the results of the evidence where hackathons have been used to overcome newcomer challenges in OSS and beyond.


\begin{table*}[!ht]
\centering

\begin{adjustbox}{angle=90}
\begin{minipage}{\textheight}
\captionof{table}{Evidence in literature where hackathons were used to overcome newcomer challenges in OSS projects}
\label{t:evidence} 

\resizebox{\textheight}{!}{%
\begin{tabular}{@{}llll@{}}
\toprule

\textbf{Category}                             & \textbf{Sub Category}                                           & \textbf{Sub Sub Category}                                                                                                                                                                                                                                                                                                     & \textbf{Classification}                \\ \midrule
\multirow{5}{*}{Finding a way to start~\citep{STEINMACHER201567}}       & Find appropriate task to start with~\citep{STEINMACHER201567,NewcomersBarriers}                             &                                                                                                                                                                                                                                                                                                                               & P-E-OTH~\citep{nolte2020support,10.1145/3502717.3532157} \\ \cmidrule(l){2-4} 
                                              & Find a mentor~\citep{STEINMACHER201567,10.1145/2675133.2675215}                                                   &                                                                                                                                                                                                                                                                                                                               & P-E-OTH~\citep{nolte2020support,10.1145/3502717.3532157} \\ \cmidrule(l){2-4} 
                                              & Difficulties locating the bug they chose to fix~\citep{7985661}                 &                                                                                                                                                                                                                                                                                                                               & P-E-OTH~\citep{nolte2020support,10.1145/3502717.3532157} \\ \cmidrule(l){2-4} 
                                              & Bug reproduction~\citep{10.1145/3125433.3125446}                                                &                                                                                                                                                                                                                                                                                                                               & P-E-OTH~\citep{nolte2020support,10.1145/3502717.3532157} \\ \cmidrule(l){2-4} 
                                              & Willingness to start with a complex task~\citep{NewcomersBarriers}                        &                                                                                                                                                                                                                                                                                                                               & P-E-OTH~\citep{nolte2020support,10.1145/3502717.3532157} \\ \midrule
\multirow{8}{*}{Technical hurdles~\citep{STEINMACHER201567}}            & \multirow{3}{*}{Issue setting up local workspace~\citep{STEINMACHER201567}}               & Problems to build the project~\citep{10.1145/3125433.3125446}                                                                                                                                                                                                                                                                                                 & P-E-OTH~\citep{10.1145/3502717.3532157,8816922,10.1145/3422392.3422479} \\ \cmidrule(l){3-4} 
                                              &                                                                 & Setup the development environment~\citep{10.1145/3125433.3125446}                                                                                                                                                                                                                                                                                             & P-E-OTH~\citep{10.1145/3502717.3532157,8816922,10.1145/3422392.3422479} \\ \cmidrule(l){3-4} 
                                              &                                                                 & Difference in the devices that mentors and mentees use~\citep{NewcomersBarriers}                                                                                                                                                                                                                                                                        & NO-E                            \\ \cmidrule(l){2-4} 
                                              & Code Complexity~\citep{STEINMACHER201567}                                                 &                                                                                                                                                                                                                                                                                                                               & P-E-OTH~\citep{10.1145/3502717.3532157,8816922,10.1145/3422392.3422479} \\ \cmidrule(l){2-4} 
                                              & Software architecture complexity~\citep{STEINMACHER201567}                                &                                                                                                                                                                                                                                                                                                                               & P-E-OTH~\citep{8816922,10.1145/3422392.3422479} \\ \cmidrule(l){2-4} 
                                              & \multirow{3}{*}{Submission process is too long and too complex~\citep{7985661}} & Submission technique~\citep{10.1145/3125433.3125446}                                                                                                                                                                                                                           & P-E-OTH~\citep{10.1145/3502717.3532157} \\ \cmidrule(l){3-4} 
                                              &                                                                 & Issue tracker complexity~\citep{10.1145/3125433.3125446}                                                                                                                                                                                                                                                                                                      & NO-E                            \\ \cmidrule(l){3-4}
                                              &                                                                 & Bureaucracy~\citep{10.1145/3125433.3125446}                                                                                                                                                                                                                                                                                       & NO-E \\ \cmidrule(l){3-4}
                                              &                                                                 & Long project processes~\citep{NewcomersBarriers}                                                                                                                                                                                                                                                                                                        & NO-E                            \\ \midrule
\multirow{4}{*}{Poorly documented code~\citep{STEINMACHER201567}}       & Too much documentation~\citep{STEINMACHER201567}                                          &                                                                                                                                                                                                                                                                                                                               & NO-E                            \\ \cmidrule(l){2-4} 
                                              & Outdated documentation~\citep{STEINMACHER201567}                                          &                                                                                                                                                                                                                                                                                                                               & NO-E                            \\ \cmidrule(l){2-4} 
                                              & Unclear code comments~\citep{STEINMACHER201567}                                           &                                                                                                                                                                                                                                                                                                                               & NO-E                            \\ \cmidrule(l){2-4} 
                                              & Lack of documentation~\citep{STEINMACHER201567}                                          &                                                                                                                                                                                                                                                                                                                               & NO-E                            \\ \midrule
\multirow{5}{*}{Newcomer previous knowledge~\citep{STEINMACHER201567}} & Lack of technical experience~\citep{STEINMACHER201567}                                    &                                                                                                                                                                                                                                                                                                                               & P-E-OTH~\citep{nolte2020support,8816922} \\ \cmidrule(l){2-4} 
                                              & Lack of domain experience~\citep{STEINMACHER201567}                                       &                                                                                                                                                                                                                                                                                                                               & P-E-OTH~\citep{nolte2020support,8816922} \\ \cmidrule(l){2-4} 
                                              & Lack of knowledge of project practices~\citep{STEINMACHER201567}                          &                                                                                                                                                                                                                                                                                                                               & P-E-OTH~\citep{nolte2020support,8816922} \\ \cmidrule(l){2-4} 
                                              & Unfamiliar project management schemes~\citep{7985661}                           &                                                                                                                                                                                                                                                                                                                               & P-E-OTH~\citep{nolte2020support,8816922} \\ \cmidrule(l){2-4} 
                                              & Lack of knowledge of the project's programming language~\citep{10.1145/2593728.2593732}         &                                                                                                                                                                                                                                                                                                                               & P-E-OTH~\citep{nolte2020support,8816922} \\ \midrule
Social Interaction~\citep{STEINMACHER201567}                            & Not receiving a (timely) answer~\citep{STEINMACHER201567,10.1145/2675133.2675215}                                 & \begin{tabular}[c]{@{}l@{}}Delayed answers~\citep{10.1145/2675133.2675215}\\ Low responsiveness~\citep{10.1145/2675133.2675215}\end{tabular}                                                                                                                                                                                                                                                  & P-E-OTH~\citep{nolte2020support,10.1145/3510309.3510324} \\ \cmidrule(l){2-4} 
                                              & Receive an improper answer~\citep{STEINMACHER201567}                                      & \begin{tabular}[c]{@{}l@{}}Impolite answers~\citep{10.1145/2675133.2675215}\\ Receiving answers with too advanced/complex context~\citep{10.1145/2675133.2675215}\end{tabular}                                                                                                                                                                                                                & P-E-OTH~\citep{nolte2020support,10.1145/3510309.3510324} \\ \cmidrule(l){2-4} 
                                              
                                              & \multirow{6}{*}{Communication barriers}                         & English level~\citep{10.1145/2675133.2675215,NewcomersBarriers}                                                                                                                                                                                                                                                                                                                 & NO-E                            \\ \cmidrule(l){3-4} 
                                              &                                                                 & Newcomer not sending meaningful messages~\citep{10.1145/2675133.2675215}                                                                                                                                                                                                                                                                                     & P-E-OTH~\citep{10.1145/3510309.3510324} \\ \cmidrule(l){3-4} 
                                              &                                                                 & Making useless comments in the mailing list~\citep{10.1145/2675133.2675215}                                                                                                                                                                                                                                                                                   & P-E-OTH~\citep{10.1145/3510309.3510324} \\ \cmidrule(l){3-4} 
                                              &                                                                 & Time zone and place barriers~\citep{NewcomersBarriers}                                                                                                                                                                                                                                                                                                  & P-E-OTH~\citep{10.1145/3510309.3510324} \\ \cmidrule(l){3-4} 
                                              &                                                                 & Lack of interpersonal skills in mentors~\citep{NewcomersBarriers}                                                                                                                                                                                                                                                                                       & P-E-OTH~\citep{10.1145/3510309.3510324} \\ \cmidrule(l){3-4}   
                                              &                                                                 & Cultural differences~\citep{NewcomersBarriers}                                                                                                                                                                                                                                                                                       & P-E-OTH~\citep{10.1145/3510309.3510324} \\ \cmidrule(l){3-4}
                                              
                                              &                                                                 & Some newcomers need to contact a real person~\citep{10.1145/2675133.2675215}                                                                                                                                                                                                                                                                                  & P-E-OTH~\citep{10.1145/3422392.3422479,10.1145/3510309.3510324} \\ \midrule
Individuals problems                          & Newcomers' personal issues                                      & \begin{tabular}[c]{@{}l@{}}Lack of clear professional goals~\citep{NewcomersBarriers}\\ Fear of judgment~\citep{NewcomersBarriers}\\ Low self-efficacy~\citep{NewcomersBarriers}\\ Performance anxiety~\citep{NewcomersBarriers}\\ Newcomer’s personality conflicts with the role~\citep{NewcomersBarriers}\\ Newcomer’s inability to improve upon criticism~\citep{NewcomersBarriers}\\ Difficulty in time-management~\citep{NewcomersBarriers}\\ Difficulty in managing different accounts~\citep{NewcomersBarriers}\\ Shyness~\citep{10.1145/2675133.2675215}\end{tabular} & NO-E                            \\ \cmidrule(l){2-4} 
                                              & \multirow{5}{*}{Mentors' issues}                                & Difficulty in time-management~\citep{NewcomersBarriers}                                                                                                                                                                                                                                                                                                 & INC-E~\citep{10.1145/3368089.3409724}                   \\ \cmidrule(l){3-4} 
                                              &                                                                 & Handling a large number of mentees~\citep{NewcomersBarriers}                                                                                                                                                                                                                                                                                            & INC-E~\citep{10.1145/3368089.3409724}                   \\ \cmidrule(l){3-4} 
                                              &                                                                & Difficulty in switching context~\citep{NewcomersBarriers}                                                                                                                                                                                                                                                                                               & INC-E~\citep{10.1145/3368089.3409724}                  \\ \cmidrule(l){3-4} 
                                              &                                                                 & Not having a formal procedure for introducing the community~\citep{NewcomersBarriers}                                                                                                                                                                                                                                                                   & NO-E                            \\ \cmidrule(l){3-4} 
                                              &                                                                 & Difficulty in managing different accounts~\citep{NewcomersBarriers}                                                                                                                                                                                                                                                                                     & NO-E                            \\ \bottomrule

\end{tabular}
}
\end{minipage}
\end{adjustbox}
\vspace{-20pt}
\end{table*}

\subsubsection{Finding a Way to Start}
\label{subsec:results:sub:challenges:findingawaytostart}
This category is related to newcomers' challenges to find the first steps to become engaged in a project and find their way to interact with the existing team of contributors \citep{STEINMACHER201567}. One problem reported by~\citet{STEINMACHER201567,NewcomersBarriers} is to \textit{\quotes{find an appropriate task to start with}}. \citet{STEINMACHER201567,10.1145/2675133.2675215} also report challenges to \textit{\quotes{find a mentor}} who guides a newcomer through onboarding and integration them into the team. \textit{\quotes{Difficulties locating bugs they choose to fix}} is a related challenge reported by~\citet{7985661,10.1145/3125433.3125446}. Also related to bugs are challenges related to \textit{\quotes{Bug reproduction}} which is reported by~\citet{10.1145/3125433.3125446}. Newcomers struggle to reproduce bugs and test if their fix resolved them. One last challenge reported by~\citet{NewcomersBarriers} is that newcomers sometimes \textit{\quotes{start with a too complex task}} for their skillset. They might struggle to understand the task and are not able to accomplish what they desire with their current experience in the OSS project.

\subsubsection{Technical Hurdles}
\label{subsec:results:sub:challenges:technicalhurdles}
This category includes technical challenges related to handling code. One of the main issues for newcomers reported by~\citet{STEINMACHER201567} is \textit{\quotes{setting up the local workspace}}. This is always a challenge for new developers joining a project. They often face problems building the project~\citep{10.1145/3125433.3125446}, setting up the development environment~\citep{10.1145/3125433.3125446}, and using different devices~\citep{NewcomersBarriers}. Other barriers reported by~\citet{STEINMACHER201567} are \textit{\quotes{Code complexity}} and \textit{\quotes{Software architecture complexity}} which always need time and effort by newcomers to get a grasp of the project and its specifics. One main issue reported by~\citet{7985661} is that the \textit{\quotes{Submission process is too long and too complex}}. This relates to issue tracker complexity~\citep{10.1145/3125433.3125446}, bureaucracy~\citep{10.1145/3125433.3125446} and submission techniques~\citep{10.1145/3125433.3125446} (ex. patch-based or pull requests) which can be different for each project.

\subsubsection{Poorly Documented Code}
\label{subsec:results:sub:challenges:poorlydocumentedcode}
This category is related to documentation, an important aspect of any software development project. \citet{STEINMACHER201567} categorized barriers related to documentation into \textit{\quotes{Too much documentation}}, \textit{\quotes{Outdated documentation}}, \textit{\quotes{Unclear code comments}}, and \textit{\quotes{Lack of documentation}}. Lack of documentation was also reported by~\citet{10.1145/3125433.3125446} referring to newcomers having to use scripts that are not fully documented and must be learned by observation or trial-and-error.

\subsubsection{Newcomers' Previous Knowledge}
\label{subsec:results:sub:challenges:previousknowledge}
This category is related to inadequate previous knowledge that could become a challenge for newcomers when joining an OSS project. \textit{\quotes{Lack of technical experience}}, \textit{\quotes{Lack of domain experience}}, and \textit{\quotes{Lack of knowledge of project practices}} are reported by~\citet{STEINMACHER201567}. Other newcomers' previous knowledge challenges are \textit{\quotes{Unfamiliar project management schemes}}~\citep{7985661} and \textit{\quotes{Lack of knowledge of the project's programming language}}~\citep{10.1145/2593728.2593732}.

\subsubsection{Social Interaction}
\label{subsec:results:sub:challenges:socialinteraction}
This category refers to one of the most common groups of challenges faced by newcomers in OSS projects. Subcategories include \textit{\quotes{Not receiving a timely answer}}~\citep{STEINMACHER201567,10.1145/2675133.2675215}, e.g., in the form of delayed answers, and \textit{\quotes{Receiving an improper answer}}~\citep{STEINMACHER201567}, e.g., in the form of impolite answers or answers with advanced or complex content. A subcategory that we derived covers \textit{Communication barriers}. It includes barriers related due to insufficient command of the English language~\citep{10.1145/2675133.2675215,NewcomersBarriers}, newcomers not sending meaningful messages~\citep{10.1145/2675133.2675215}, making useless comments in mailing lists~\citep{10.1145/2675133.2675215}, issues related to timezone and geographical location~\citep{NewcomersBarriers}, lack of interpersonal skills of mentors~\citep{NewcomersBarriers}, cultural differences~\citep{NewcomersBarriers}, and the need of newcomers to contact a person face-to-face~\citep{10.1145/2675133.2675215}. 

\subsubsection{Individuals Problems}
\label{subsec:results:sub:challenges:individualsproblems}
This category was derived by us, and it consists of various problems of individuals in OSS projects, arising either from newcomers themselves or from mentors, and affecting the newcomers' contributions. 

The first subcategory contains \textit{Newcomers' personal issues} that could affect their integration in an OSS project, including lack of clear professional goals~\citep{NewcomersBarriers}, fear of judgement~\citep{NewcomersBarriers}, low self-efficacy~\citep{NewcomersBarriers}, performance anxiety~\citep{NewcomersBarriers}, newcomer's personality conflicts with the role~\citep{NewcomersBarriers}, newcomer's inability to improve upon criticism~\citep{NewcomersBarriers}, difficulty in time management~\citep{NewcomersBarriers}, difficulty in managing different accounts~\citep{NewcomersBarriers}, and shyness~\citep{10.1145/2675133.2675215}.

The second subcategory contains \textit{Mentor issues} that could become barriers for newcomers in OSS projects. Mentors are often needed by newcomers to integrate into the team quickly. Mentor issues relate to difficulties with time-management~\citep{NewcomersBarriers}, handling a large number of mentees~\citep{NewcomersBarriers}, switching context~\citep{NewcomersBarriers}, not having a formal procedure for introducing the community~\citep{NewcomersBarriers}, and managing different accounts~\citep{NewcomersBarriers}.


\subsection{Evidence on How Hackathons Address Newcomer Challenges in OSS Projects}
\label{subsec:results:evidence}
In this subsection, we present evidence on how hackathons have addressed newcomers' challenges in OSS projects and beyond (see the right-most column of Table~\ref{t:evidence}).

\subsubsection{Finding a Way to Start}
\label{subsec:results:sub:evidence:findingawaytostart}
This category is related to newcomer challenges while becoming engaged with the project and finding their way to integrate with the existing team. The challenges in this category are directly related to learning and coaching when starting participation in a project. This matches well with hackathons set-up as time-bounded events focusing on participants as potential project newcomers, with assigned mentors for each team. 
\citet{nolte2020support} reports about supporting newcomers in scientific hackathons and shares recommendations about how mentors should focus on mentoring the team, and their learning rather than the project completion. They also found that teams taking ownership of their projects, receiving proper support from mentors, and receiving learning-oriented support reported positive outcomes. Besides, \citet{10.1145/3502717.3532157} also found positive outcomes for learning through hackathons. They used mentors to support teams, guide them, and carefully structure the hackathon events in order to improve their authentic learning.

\subsubsection{Technical Hurdles}
\label{subsec:results:sub:evidence:technicalhurdles}
This category refers to technical challenges related to the handling of program code by newcomers. Similar to the first category, newcomers' challenges related to building the project, setting up the development environment, code complexity, submission technique, and software architecture complexity can also be solved by learning~\citep{10.1145/3502717.3532157} and coaching by mentors. Researchers~\citet{10.1145/3422392.3422479} used hackathons for learning and to engage students to learn and adapt software engineering practices, highlighting positive outcomes when hackathon participants learn from their peers, share knowledge and simply have an experience different from that provided in a classroom setting. On the other hand, we could not find studies that specifically mention challenges like issue tracker complexity, bureaucracy, long project processes, and difference in the devices that mentors and mentees use.

\subsubsection{Poorly Documented Code}
\label{subsec:results:sub:evidence:poorlydocumentedcode}
This category relates to documentation, which is an important aspect of any software development project. We could not find any evidence that hackathons were used to overcome any of the subcategories related to poorly documented code, including \textit{\quotes{too much documentation}}, \textit{\quotes{outdated documentation}}, \textit{\quotes{unclear code comments}}, and \textit{\quotes{lack of documentation}}. We found in one existing SLR~\citep{medina2020what} that documentation could be one outcome of a hackathon, implying that hackathons may be used to improve the documentation or generate new documentation and, thus, OSS projects might use hackathons for that goal. However, this finding is not related to our focus on helping newcomers integrate faster in an OSS project. 

\subsubsection{Newcomers' Previous Knowledge}
\label{subsec:results:sub:evidence:previousknowledge}
Newcomers' challenges are related to the lack of knowledge about the project as a whole or aspects such as project management, domain, or programming language. The usage of hackathons for improving the learning curve is discussed by~\citet{10.1145/3422392.3422479} and positive evidence has been achieved by allowing participants to interact and talk with mentors and stakeholders. Other researchers~\citet{8816922} report on improving learning skills through datathon events. Researchers~\citet{nolte2020support} also provided several recommendations for choosing experienced mentors with previous knowledge about the domain, community, and project practices that can help the hackathon participants to achieve their goal and improve their engagement.

\subsubsection{Social Interaction}
\label{subsec:results:sub:evidence:socialinteraction}
Hackathons are mainly social events where participants interact and work together to achieve a goal set during the event, so it can be used to overcome most of the challenges in the category. Challenges related to \textit{\quotes{Not receiving a (timely) answer}} and \textit{\quotes{Receive an improper answer}} are mainly discussed by~\citet{nolte2020support} suggesting that assigning a mentor with previous knowledge about the community may help with guiding the team towards the idea and the solution and eliminating wrong communications between newcomers and mentors. Researchers~\citet{10.1145/3510309.3510324} found positive evidence for hackathons improving social interactions and communication which also covers the challenges faced by newcomers related to \textit{Communication barriers} - except the English level barrier for which we were not able to find any evidence of hackathons being used to solve it.

\subsubsection{Individuals Problems}
\label{subsec:results:sub:evidence:individualproblems}
This category is connected to the individual problems in the project that could arise from the newcomers or from the mentors. Few of the challenges in this category related to \textit{\textbf{Mentors' issues}} are mentioned in the literature, however, no conclusive evidence was mentioned. Researchers~\citet{10.1145/3368089.3409724} recommend assigning 2 mentors (one experienced and another new mentor) can help in balancing the load on the assigned mentor, thus can help to overcome mentors' challenges like difficulty in time-management, handling a large number of mentees, and difficulty in switching context. No evidence was found of the usage of hackathons to overcome other challenges in this category related to newcomers or mentors.



\section{\uppercase{Limitations and Threats to Validity}}
\label{sec:limitations}
Our findings are based on a literature search utilizing specific search terms. This approach does not guarantee that we found all relevant papers. Moreover, our selection of digital libraries might not have been comprehensive. 

Since we only used refereed research publications and did not include gray literature, we might have missed challenges and barriers and evidence -- both positive and negative -- of using hackathons to address them. Moreover, the data extraction from the relevant literature as well as the analysis of evidence is subject to interpretation bias.

\section{\uppercase{Conclusion and future work}}
\label{sec:conclusion}
Based on our findings, several of the newcomers' challenges and barriers might be solved by conducting hackathons and there is potential in using hackathons to overcome newcomers' challenges in OSS projects, allowing them to integrate faster into the project. 

The work presented in this paper is the first step in a larger research undertaking. It should thus be perceived as a position paper paving the ground for future empirical work on how hackathons can support newcomers to join and become productive members of OSS projects.

\section*{\uppercase{Acknowledgements}}
\label{sec:acknowledgment}
Part of this work was funded by grant PRG1226 of the Estonian Research Council.

\bibliographystyle{apalike}
{\small
\bibliography{example}}

\end{document}